\documentclass[11pt]{article}
\usepackage{geometry}
\usepackage{graphicx}
\usepackage[section]{placeins}
\usepackage{microtype}
\usepackage{amssymb}
\usepackage{amsmath}
\usepackage{epstopdf}
\usepackage{cite}
\usepackage{bbm}
\usepackage{multirow}
\usepackage{enumerate}
\usepackage{ulem}
\usepackage{xcolor}
\usepackage{appendix}
\usepackage[hyperindex=true,
          pdfstartview=FitH,
          bookmarksnumbered=true,
          bookmarksopen=true,
          citecolor=blue,
          linkcolor=blue,
          colorlinks=true,
          pdfborder={0 0 0}]{hyperref}

\newcommand{\ba}{\begin{align}}
\newcommand{\ea}{\end{align}}
\newcommand{\be}{\begin{equation}}
\newcommand{\en}{\end{equation}}
\newcommand{\bea}{\begin{eqnarray}}
\newcommand{\ena}{\end{eqnarray}}

\title{Transient resonances of chaotic orbits in magnetized black hole spacetimes: an event-based frequency-ratio diagnostic}

\author{
Wenfu Cao$^{1}$\footnote{202411100001@stu.ujn.edu.cn},
Ying Wang$^{2}$\footnote{wangying424524@163.com},
Hongsheng Zhang$^{1}$\footnote{sps\_zhanghs@ujn.edu.cn (corresponding author)}
\\[0.8em]
\small
$^{1}$School of Physics and Technology, University of Jinan,\\
\small
336 West Road of Nan Xinzhuang, Jinan, Shandong 250022, China
\\[0.4em]
\small
$^{2}$School of Mathematics, Physics and Statistics,\\
\small
Shanghai University of Engineering Science, Shanghai 201620, China
}

\begin{document}

\maketitle

\begin{abstract}
The geodesic equations of the Kerr spacetime are separable, and this separability is what makes orbital dynamics around black holes tractable. An external electromagnetic field removes it, the motion turns chaotic, and the orbit wanders between resonances for irregular and unequal intervals; a long-time Poincar\'e section returns a single static blur, so that it records where an orbit has been but not when. We introduce the sliding-window frequency ratio, an event-based diagnostic that counts polar turning events within a window of fixed radial-cycle number and returns the local ratio of the two frequencies without any spectral decomposition; for regular motion with bounded count deviations, the error falls off as the window widens. We first establish  that a single globally chaotic orbit in a magnetized black hole spacetime pays temporally ordered visits to distinct resonances, and that the corresponding section structure appears only in those intervals that the event counts select. A charged orbit in magnetized Kerr spacetime dwells near 3/5 and later near 4/7, opening fivefold and sevenfold structures, while a photon in Schwarzschild--Melvin spacetime holds 4/5 across a fivefold pattern. Because we keep every event time, we can replot each selected interval as its own section, and integer counts then agree with phase-space geometry. These visits control the phase accumulated along an orbit, and they therefore bear on any orbit-based probe of strong-field black hole geometry.
\end{abstract}

\section{Introduction}\label{sec:introduction}

Particle motion near black holes provides a setting for studying
integrability, resonances, and chaos. In vacuum Kerr spacetime, an additional
integral of motion makes the geodesic equations separable~\cite{Carter:1968rr}, and
bound orbits can be described by radial, polar, and azimuthal fundamental
frequencies~\cite{Schmidt:2002frequencies}. A radial--polar resonance occurs
when these two fundamental frequencies of the same particle have a rational
ratio. The locations of these resonances and their dependence on orbital
parameters have been studied in Kerr spacetime~\cite{Brink:2015resonances}.
An external electromagnetic
field can break integrability for charged particles, allowing resonant
island chains and chaotic regions to appear in phase
space~\cite{Mukherjee:2023resonance}.

\begin{figure}[!t]
\centering
\includegraphics[width=0.6\textwidth]{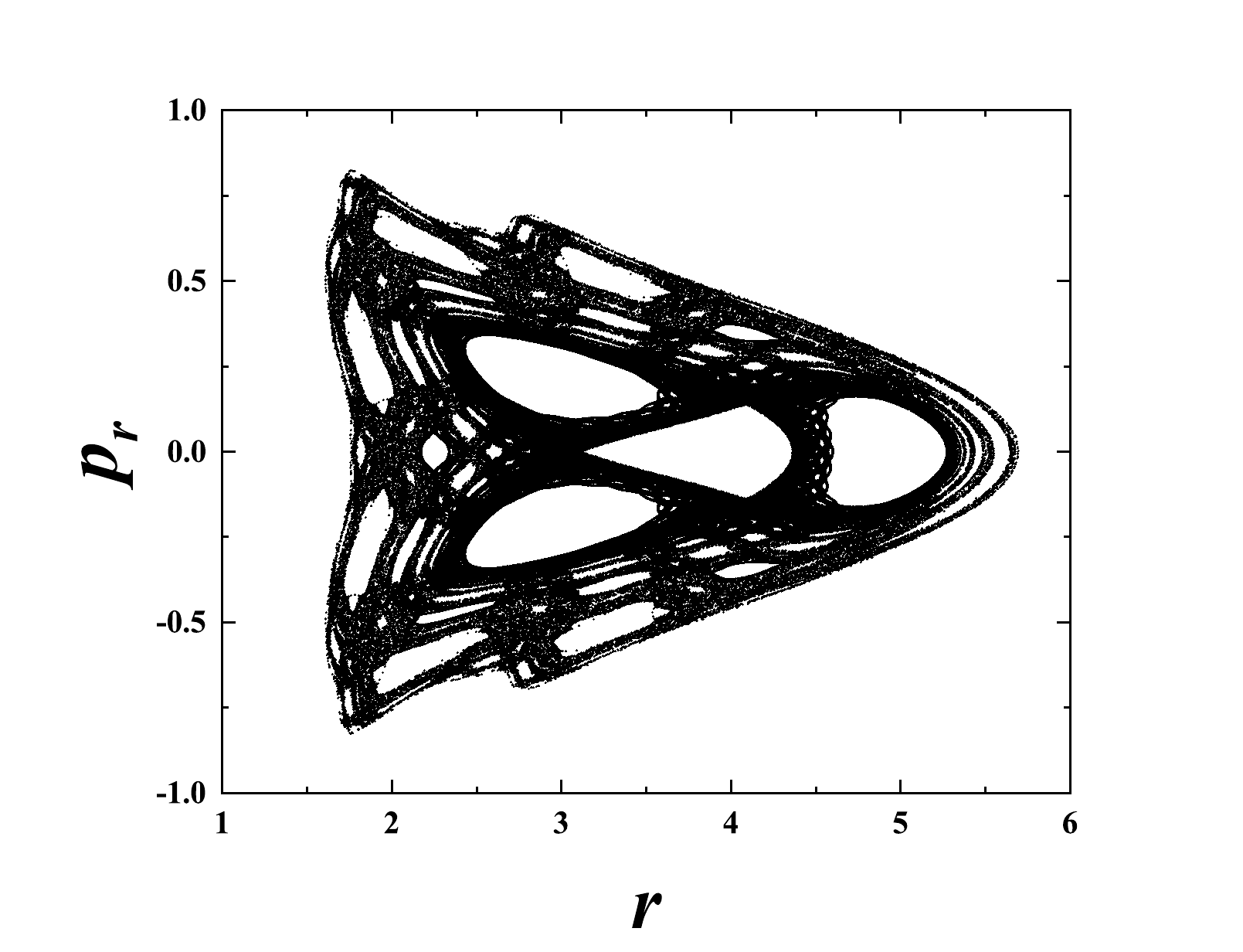}
\caption{Long-time Poincar\'e section of the chaotic charged-particle orbit
studied in Sec.~\ref{sec:resonance_geometry}, with $E=0.905$, $L=2$,
$a=0.99$, $b=0.105$, and $r_0=4.6$, recorded at $\theta=\pi/2$ with
$\dot\theta>0$ over $0\leq\tau\leq10^7$ in the $(r,p_r)$ plane. The same
section appears in the lower-right panel of Fig.~\ref{fig:charged_sections}.}
\label{fig:long_time_section}
\end{figure}

In phase space where regular and chaotic regions coexist, a chaotic orbit
can show different local patterns of motion at different times. It may
remain near a resonant structure for a long time before leaving to explore
a wider region, a phenomenon known as
stickiness~\cite{Contopoulos:2010stickiness}. A Poincar\'e section shows the
geometric distribution, but a section accumulated over a long integration
superposes points from different times. The complex distribution in
Fig.~\ref{fig:long_time_section} does not directly show which structures
the orbit approaches in sequence or how long each visit lasts. Following
this evolution requires retaining the time information and relating the
oscillatory motion in each interval to its section structure.

Frequency analysis is widely used to study orbital structure. Fundamental
frequencies and their variation can be used to identify regular motion,
resonances, and chaotic diffusion~\cite{Laskar:1993frequency}.
Time--frequency analysis also shows when frequency changes occur along a
chaotic trajectory. It can track temporary residence near resonances and
transitions between them, allowing comparison with motion in phase
space~\cite{Chandre:2003timefrequency}. The rotation number describes
frequency relations through the rotational motion of an orbit.
In a common Poincar\'e-section implementation, it is estimated
by accumulating the angles between consecutive section points relative
to a reference center. With a suitable section and angle convention, it
can be related to the orbital frequency ratio. Varying the initial
conditions gives a rotation curve, on which plateaus at rational values
can identify resonant island chains. This method has been used in studies of relativistic orbital dynamics,
including non-Kerr spacetimes and charged-particle motion around a Kerr black
hole with an external magnetic
field~\cite{LukesGerakopoulos:2010signature,Mukherjee:2023resonance,LukesGerakopoulos:2012zipoy,Zelenka:2020spin,Destounis:2020symmetry,Destounis:2021glitches,Deich:2022quadratic,EduarteRojas:2022kerrlike,Chen:2022eob,Destounis:2023boson,Shahzadi:2023hartlethorne,Destounis:2026environment,Strateny:2026matter,Yuan:2026spin}.

Finite-time rotation information has also been used to study local
structures in chaotic motion. The distribution of finite-time rotation
numbers over initial conditions can reveal transport
structures~\cite{Szezech:2013ftrn}. Changes in rotation-number calculations
have also been used to detect sticky motion and estimate trapping times
in the standard map~\cite{Santos:2019sticky}. In relativistic orbital
dynamics, Fourier analysis over successive time intervals of a non-Kerr
inspiral driven by radiation reaction has revealed frequency-ratio
plateaus associated with resonance
crossing~\cite{Apostolatos:2009nonKerr}. For conservative motion, power
spectra and corresponding Poincar\'e sections have been used to compare
different segments of individual geodesics in the field of a black hole
surrounded by a disc. These comparisons distinguish sticky phases from
chaotic motion spread over a wider region~\cite{Semerak:2012geodesics}.
Such results provide a basis for studying how local structures change
along an orbit. Here we study conservative chaotic
orbits at fixed parameters, focusing on their temporary residence near
different resonances and the local frequency relations during these
stages.

Oscillation frequencies can also be estimated from recurring events along
an orbit. Zero crossings, for example, have been used to estimate orbital
fundamental frequencies and construct frequency
maps~\cite{Yurin:2024zerocrossing}. For orbits with well-defined radial and
polar oscillations, turning events in a specified direction provide a
direct way to count oscillation cycles. We use these events to construct
the sliding-window frequency ratio (SWFR). We record one selected turning
event per complete oscillation in each coordinate. Each window spans a
fixed number of radial cycles; the polar events within it are counted,
and the radial cycle count is divided by the polar event count to estimate
the frequency ratio. Moving the window along the orbit follows changes
in this local oscillation relation. The events and their occurrence times
can be recorded during integration, and the same sequence can be reused
for different window widths. This calculation requires neither spectral
extraction nor a reference center on the section. Its applicability
requires a clear correspondence between the selected events and complete
oscillation cycles.

We treat intervals in which SWFR remains near a low-order rational value
as resonance candidates. The event-time table links these intervals to
orbit segments, whose Poincar\'e sections are then computed over the same
times. The frequency relation locates a candidate stage, while the
section provides geometric evidence in phase space. Agreement between
the two supports temporary residence near the associated resonance. For
a globally chaotic orbit, this analysis concerns local motion over a
finite interval; the frequency plateau and its section must be interpreted
together.

We first test SWFR on integrable Kerr orbits with known frequency ratios
and examine how the window width affects counting fluctuations. We then
apply it to chaotic charged-particle orbits in an external magnetic
field. In the main example, intervals near $3/5$ and $4/7$ along one orbit
have fivefold and sevenfold local section structures, respectively.
Sections for consecutive time intervals show these stages and the
evolution between them. A second orbit illustrates an interval near
$1/2$ and its two-lobed structure. Finally, we study photon orbits in
Schwarzschild--Melvin spacetime, where the magnetic field affects the
geometry through its own gravity and photons follow the resulting null
geodesics~\cite{LimaJunior:2021shadows}. This example tests the
applicability of SWFR in a different dynamical setting.

Section~\ref{sec:framework} defines SWFR and discusses its frequency
interpretation, finite-window errors, and relation to resonant section
structures. Section~\ref{sec:kerr_benchmark} presents the integrable Kerr
tests. Sections~\ref{sec:charged_kerr} and \ref{sec:melvin} examine the
charged-particle and photon orbits, respectively.
Section~\ref{sec:conclusions} summarizes the results and discusses the
method's range of applicability.

\section{Sliding-window frequency ratio and resonance structure}\label{sec:framework}

When radial and polar turning events are well defined, their counts describe
the relative oscillation rates. We use this relation to define SWFR and
examine its frequency interpretation, finite-window errors, and connection
to periodic structures on Poincar\'e sections.

\subsection{Turning events and the definition of SWFR}\label{sec:events}

Consider an orbit that oscillates in both the radial and polar coordinates.
Let $s$ be an evolution parameter that increases monotonically along the
orbit. For massive particles, $s$ is proper time; for photons, it is an affine
parameter.
We select one type of turning event in each coordinate and use the event
counts within a finite window to estimate the local frequency ratio.

For radial motion, we select turning points where the radial momentum
$p_r$ changes from negative to positive, as the orbit turns from inward to
outward motion. This gives one event per complete radial oscillation.
The event times are
\[
s_0,s_1,\ldots,s_m,\ldots,s_M,
\]
where $s_0$ is the first recorded radial event time and $M$ is the index of
the last event. Successive events at $s_m$ and $s_{m+1}$ delimit one radial
cycle, so $m$ counts the radial cycles completed since $s_0$.

For polar motion, we select turning points where the polar momentum
$p_\theta$ changes from negative to positive, again giving one event per
complete polar oscillation. With $s_0$ as the reference time, let $C_m$ be the
cumulative number of polar events in $[s_0,s_m)$, with $C_0=0$. All counting
intervals include the left endpoint and exclude the right endpoint. A polar
event coinciding with a radial event therefore belongs to the radial cycle
beginning at that time.

Choose a positive even integer $W$ as the window width. Here $W$ counts
complete radial cycles, not integration steps or sampled trajectory points.
Starting at radial event $k$, we take the next $W$ radial cycles, ending at
event $k+W$. The corresponding time interval is
\begin{equation}
[s_k,s_{k+W}).
\label{eq:swfr_window}
\end{equation}
The radial cycle count in this window is exactly $W$. The polar event count
is obtained by subtracting the cumulative counts at its endpoints:
\begin{equation}
\Delta C_W(k)=C_{k+W}-C_k.
\label{eq:polar_window_count}
\end{equation}
We assign the result for this window to the central radial event index
\begin{equation}
m_c=k+\frac{W}{2}.
\label{eq:window_center}
\end{equation}
The sliding-window frequency ratio is the radial cycle count divided by
the polar event count over the same time interval:
\begin{equation}
\rho_W\!\left(k+\frac{W}{2}\right)
=\frac{W}{\Delta C_W(k)}
=\frac{W}{C_{k+W}-C_k}.
\label{eq:swfr}
\end{equation}
We also write this quantity as $\rho_W(m_c)$. Thus $k$ specifies where the
window starts, while $m_c$ specifies where its value is plotted. Choosing
an even $W$ makes $m_c$ an integer. The ratio is defined for
$0\leq k\leq M-W$ when $\Delta C_W(k)>0$. At fixed $W$, we advance $k$ by
one radial event at a time and repeat the calculation to obtain the SWFR
curve.

For example, if a window of $W=3000$ radial cycles contains $3750$ polar
events, then $\rho_{3000}(k+1500)=3000/3750=4/5$. This value is plotted at
$m_c=k+1500$, the center of that window in radial-cycle index.

Both types of event are counted over the same interval, so their count ratio
equals the ratio of their average event frequencies. We use $\rho_W(m_c)$ to
estimate the local frequency ratio near the central event. The estimate can
represent a local stage when the relation between radial and polar
oscillations varies little within the window. A larger window averages over
more cycles but smooths more rapid changes.

We retain the event table
\[
(m,C_m,s_m),
\]
which links each event index to its cumulative count and occurrence time.
The window is centered at $m_c$ in radial-cycle index; the associated time is
$s_{m_c}$, read directly from the table. We can thus map an interval of central
event indices on the SWFR curve to a time interval along the same trajectory
and examine its phase-space structure.

\subsection{Frequency interpretation and finite-window errors}\label{sec:finite_window}

With the event-count convention of Sec.~\ref{sec:events}, the window
$[s_k,s_{k+W})$ contains $W$ complete radial cycles and $\Delta C_W(k)$ polar
events. Its duration is
\begin{equation}
\Delta s_W=s_{k+W}-s_k.
\label{eq:window_duration}
\end{equation}
The mean radial period is the total duration divided by the number of radial
cycles,
\begin{equation}
\overline T_{r,W}=\frac{\Delta s_W}{W}.
\label{eq:mean_radial_period}
\end{equation}
Since one selected event is also recorded per complete polar oscillation, we
estimate the mean polar period as
\begin{equation}
\widehat T_{\theta,W}=\frac{\Delta s_W}{\Delta C_W(k)}.
\label{eq:estimated_polar_period}
\end{equation}
The hat denotes an estimate: the window endpoints generally do not coincide
with polar events, so the window may include incomplete polar cycles at its
ends.

The relation $\Omega=2\pi/T$ between angular frequency and period gives
\begin{equation}
\widehat\Omega_{r,W}=\frac{2\pi}{\overline T_{r,W}},
\qquad
\widehat\Omega_{\theta,W}=\frac{2\pi}{\widehat T_{\theta,W}}.
\label{eq:window_frequency_estimates}
\end{equation}
Taking their ratio cancels both the common factor $2\pi$ and the window
duration, giving
\begin{equation}
\frac{\widehat\Omega_{r,W}}{\widehat\Omega_{\theta,W}}
=\frac{\widehat T_{\theta,W}}{\overline T_{r,W}}
=\frac{W}{\Delta C_W(k)}
=\rho_W(m_c).
\label{eq:swfr_frequency_interpretation}
\end{equation}
SWFR therefore estimates the ratio of the mean radial and polar angular
frequencies over the same window. Ref.~\cite{Cao:2026tpcd} discusses the
conditions that relate event counts to fundamental frequencies and tests
this relation using integrable Kerr orbits.

To estimate the counting error from a finite window, consider regular motion
with well-defined fundamental frequencies and theoretical ratio
\begin{equation}
\rho_*=\frac{\Omega_r}{\Omega_\theta}.
\label{eq:theoretical_frequency_ratio}
\end{equation}
On average, $W$ radial cycles correspond to $W/\rho_*$ polar cycles. The
integer event count can differ from this value, so we write
\begin{equation}
\Delta C_W(k)=\frac{W}{\rho_*}+\delta_W(k),
\label{eq:window_count_deviation}
\end{equation}
where $\delta_W(k)$ is the deviation of the count from the mean frequency
relation.

For two oscillations with fixed periods in a common evolution parameter and
one selected event per cycle, the window endpoints cause a count deviation
of less than one event in magnitude. More generally, the error estimate
requires a uniform bound
\begin{equation}
|\delta_W(k)|\leq D,
\label{eq:bounded_count_deviation}
\end{equation}
where $D$ may depend on the orbit and event definition, but is independent of
the window starting index and width. For fixed periods, $D=1$ suffices.
Regular orbits with the appropriate event--phase correspondence also have
bounded count deviations under the conditions in Secs.~2.2 and 2.3 of
Ref.~\cite{Cao:2026tpcd}, with a bound that may depend on the orbit.

Substituting Eq.~\eqref{eq:window_count_deviation} into the SWFR definition
in Eq.~\eqref{eq:swfr} gives
\begin{equation}
\rho_W(m_c)
=\frac{W}{W/\rho_*+\delta_W(k)}
=\frac{\rho_*W}{W+\rho_*\delta_W(k)}.
\label{eq:swfr_with_count_deviation}
\end{equation}
The difference from the theoretical frequency ratio is therefore
\begin{equation}
\rho_W(m_c)-\rho_*
=-\frac{\rho_*^2\delta_W(k)}{W+\rho_*\delta_W(k)}.
\label{eq:swfr_error_identity}
\end{equation}
For $W>\rho_*D$, the denominator is positive. Applying
Eq.~\eqref{eq:bounded_count_deviation} then yields
\begin{equation}
\left|\rho_W(m_c)-\rho_*\right|
\leq\frac{\rho_*^2D}{W-\rho_*D}
=O(W^{-1}).
\label{eq:swfr_finite_window_bound}
\end{equation}
Under the uniform bounded-deviation condition, the frequency-ratio error
bound decreases as $1/W$: the bounded count deviation contributes a smaller
fraction of the total as the number of cycles increases. This estimate
assumes an ideal event sequence and excludes errors from numerical
integration and event location.

For a chaotic orbit, SWFR also reflects changes in the local oscillation
relation. A window contained within a stage where this relation varies
little can give a representative local ratio. A window spanning different
stages mixes their contributions and smooths the transition. This averaging
effect is not described by the error bound for a fixed $\rho_*$.

The window width must balance counting accuracy and time resolution. Larger
$W$ can reduce finite-count errors during stable stages, while smaller $W$
can resolve shorter local changes. The integrable Kerr benchmark below tests
the frequency-ratio errors at different window widths. We then use a fixed
width to study local frequency structure along chaotic orbits.

\subsection{Rational frequency ratios and periodic structure on Poincar\'e sections}\label{sec:periodic_structure}

We record $(r,p_r)$ on the equatorial Poincar\'e section whenever the orbit
crosses the equatorial plane with increasing polar angle. For orbits with exactly one such
crossing per complete polar cycle, section returns count polar cycles.
This convention relates the radial-to-polar frequency ratio to the periodic
structure on the section. It has also been used to study resonances of
charged particles in Kerr spacetime with an external magnetic
field~\cite{Mukherjee:2023resonance}.

In an integrable system, a nonresonant quasiperiodic orbit traces an invariant
curve on the section as the integration time increases, whereas a resonant
periodic orbit gives a finite set of repeating points. If the theoretical
frequency ratio is $\rho_*=p/q$, with coprime positive integers $p$ and $q$,
the radial motion completes $p$ cycles while the polar motion completes $q$
cycles, and the two oscillations repeat together. With one section point per
polar cycle, this gives $q$ periodic points. The ratios $3/5$ and $4/7$ thus
correspond to period-five and period-seven structures, respectively. This
periodicity refers to the reduced radial--polar motion and does not require
the azimuthal angle to return to its initial
value~\cite{LukesGerakopoulos:2010signature}.

In a nearly integrable system satisfying the conditions of the
Poincar\'e--Birkhoff theorem, a resonant torus can break up into stable and
unstable periodic orbits. A stable orbit of period $q$ has $q$ periodic points
on the section, and regular motion around them can form a chain of $q$
islands. Under the section convention above, the denominator of the rational
frequency ratio is therefore related to the number of islands in the chain;
their shapes and widths also depend on the model and orbital parameters.
Discrete periodic points in an integrable system are distinct from stable
islands of finite width in a nonintegrable
system~\cite{Mukherjee:2023resonance}.

A chaotic orbit may remain for a finite time in the chaotic region around
an island chain. Its section points can then concentrate in organized
patterns before spreading into a larger part of phase space. Such stickiness
is related to restricted transport near stable islands and to the associated
unstable manifolds~\cite{Contopoulos:2010stickiness}. Different time segments
of the same chaotic orbit can therefore show different degrees of order on
the section, without the orbit becoming strictly periodic.

When SWFR remains near $p/q$ over an interval, we examine whether the section
over the same time interval shows a local pattern consistent with a $q$-fold
structure under this section convention. Candidate intervals near $3/5$ and
$4/7$ are therefore checked for fivefold and sevenfold structures,
respectively. Agreement between the frequency ratio and section geometry
supports a finite-time visit or episode of sticking near the resonance. The
local point sets of a chaotic orbit are not the stable islands themselves.

\section{Benchmarks with integrable Kerr orbits}\label{sec:kerr_benchmark}

We test SWFR on two integrable Kerr orbits with known fundamental frequency
ratios: one quasiperiodic orbit and one orbit with periodic radial--polar
motion. Comparing different window widths shows how finite event counts
affect the estimates of these constant ratios.

We use geometric units $G=c=M=1$, set the particle's rest mass to unity, and
take $E=0.98$, $L=2$, and $a=0.99$, with no external magnetic field.
Both orbits start at periastron with $\theta_0=\pi/2$, $\phi_0=0$, and
$p_{r0}=0$. The mass-shell constraint determines the positive initial polar
momentum $p_{\theta0}$. Table~\ref{tab:kerr_benchmark} lists the initial radii
and theoretical frequency ratios from Ref.~\cite{Cao:2026tpcd}. These ratios
were computed independently from the radial and polar periods.

\begin{table}[!htbp]
\centering
\caption{Initial radii and theoretical frequency ratios for the integrable
Kerr benchmarks. Both orbits have $E=0.98$, $L=2$, and $a=0.99$, with no
external magnetic field.}
\label{tab:kerr_benchmark}
\begingroup
\renewcommand{\arraystretch}{1.15}
\begin{tabular}{lcc}
\hline
Orbit & Initial radius $r_0$ & Theoretical ratio $\rho_*$ \\
\hline
Quasiperiodic & $4.334774218362792$ & $1/\sqrt{2}$ \\
Radial--polar periodic & $7.377636403973950$ & $4/5$ \\
\hline
\end{tabular}
\endgroup
\end{table}

We integrate the equations of motion with an eighth-order Runge--Kutta method,
using a fixed proper-time step $h=0.1$ up to $\tau=10^7$. Turning events are
recorded as defined in Sec.~\ref{sec:events}. For each orbit, we calculate
SWFR with $W=50$, $100$, $1000$, and $3000$ radial cycles from the same event
sequence, so differences between the curves arise solely from the window
width.

\begin{figure}[!htbp]
\centering
\begin{minipage}[t]{0.49\textwidth}
\centering
\includegraphics[width=\linewidth]{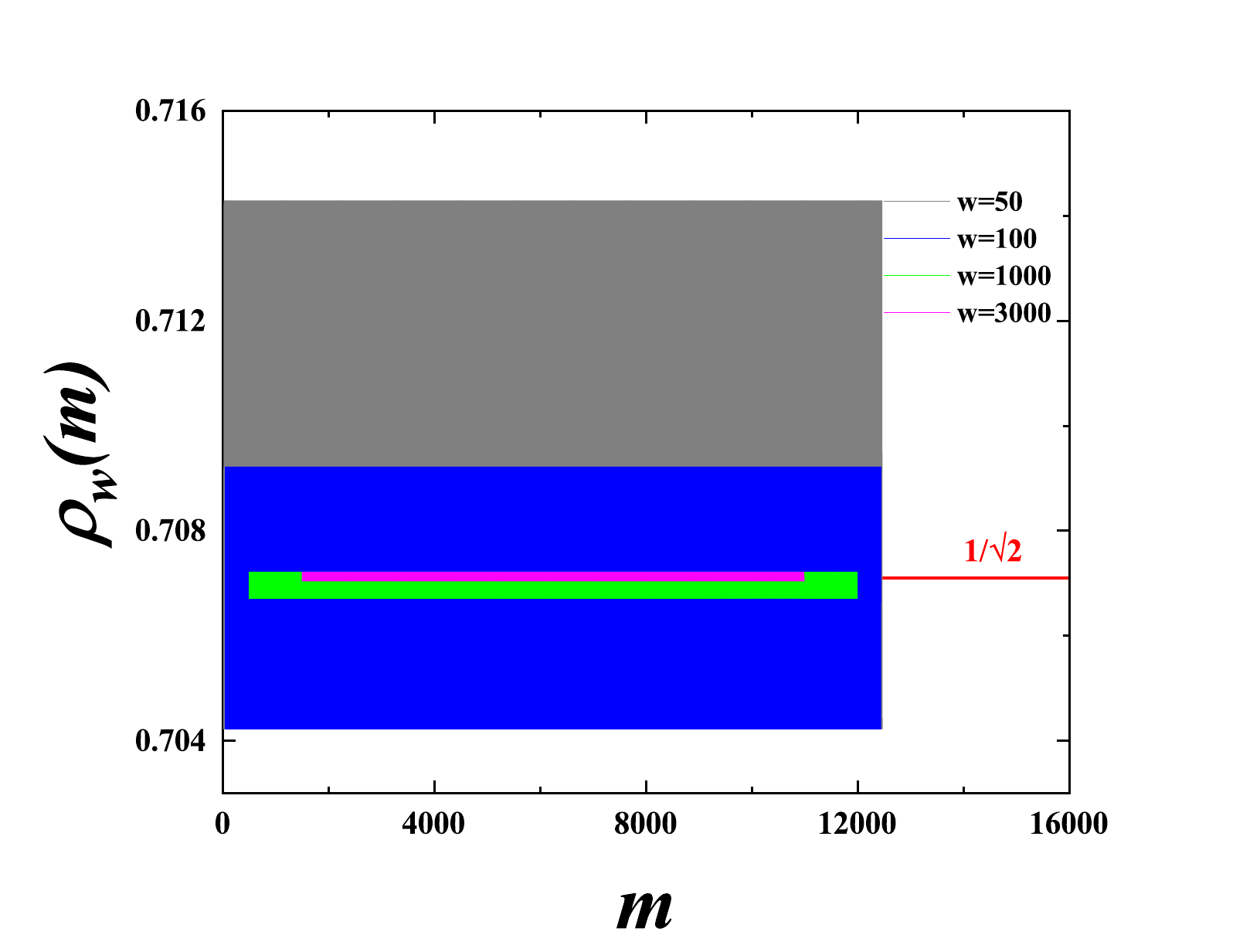}\\[-0.3em]
(a)
\end{minipage}\hfill
\begin{minipage}[t]{0.49\textwidth}
\centering
\includegraphics[width=\linewidth]{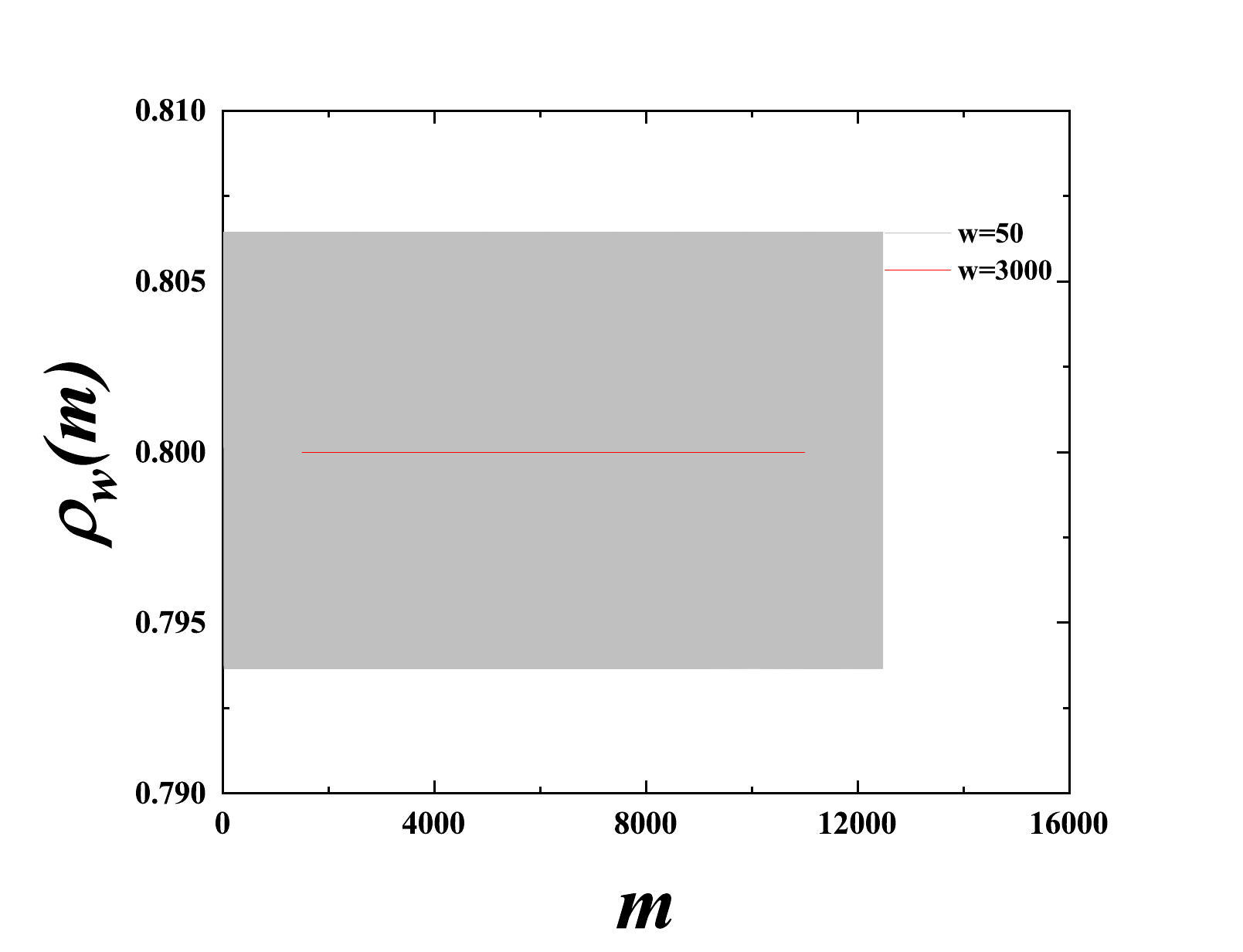}\\[-0.3em]
(b)
\end{minipage}
\caption{Dependence of SWFR on window width for two integrable Kerr orbits.
(a) A quasiperiodic orbit with $\Omega_r/\Omega_\theta=1/\sqrt{2}$.
(b) An orbit with periodic radial--polar motion and $\Omega_r/\Omega_\theta=4/5$.
The window widths are indicated in the legends. The red horizontal line
in (a) marks $1/\sqrt{2}$, and $m$ is the central radial event index.
Orbital parameters and initial radii are given in Table~\ref{tab:kerr_benchmark}.}
\label{fig:kerr_windows}
\end{figure}

For the quasiperiodic orbit in Fig.~\ref{fig:kerr_windows}(a), SWFR remains
near $1/\sqrt{2}$ without sustained drift. Small windows produce discrete
fluctuations. As $W$ increases from $50$ to $3000$, these fluctuations narrow
and the curves approach the theoretical line on the scale of the plot.
This agrees with the finite-window analysis in Sec.~\ref{sec:finite_window}:
the endpoint count deviation contributes less to the ratio as the window
includes more cycles. A ratio of finite integer counts cannot equal the
irrational value $1/\sqrt{2}$ exactly, so counting fluctuations can remain
even when the fundamental frequency ratio is constant.

For the orbit with periodic radial--polar motion and frequency ratio $4/5$
in Fig.~\ref{fig:kerr_windows}(b), four radial cycles correspond to five polar
cycles. The event-count pattern repeats, so a window containing an integer
number of repeating units can give a constant ratio. In particular,
$W=3000$ contains $750$ units and $3750$ polar events, giving
\begin{equation}
\rho_{3000}=\frac{3000}{3750}=\frac45.
\label{eq:kerr_periodic_window}
\end{equation}
The numerical data give this value at every valid window center. The windows
$W=100$ and $W=1000$ also contain complete repeating units and give the same
constant value of $0.8$.

For $W=50$, the window width is not a multiple of four. The mean frequency
relation gives $62.5$ polar cycles, but the event count must be an integer.
As the window moves, the count switches between $62$ and $63$, and SWFR
switches between $50/62\simeq0.80645$ and $50/63\simeq0.79365$.
These fluctuations arise from finite-window counting. Even for periodic
radial--polar motion, an exactly horizontal SWFR curve depends on whether
the window width matches the repeating cycle.

The benchmarks show how window width affects SWFR for regular motion.
Larger windows reduce counting fluctuations in the quasiperiodic orbit;
windows that contain complete repeating units give a constant ratio for
the orbit with periodic radial--polar motion. Since $W=3000$ substantially
suppresses finite-count fluctuations in both cases, we use it in the
nonintegrable examples below. We then compare changes in SWFR with the
Poincar\'e sections over the corresponding time intervals.

\section{Local resonance structure of charged particles in Kerr spacetime with an external magnetic field}\label{sec:charged_kerr}

We consider charged test particles around a Kerr black hole in an
asymptotically uniform magnetic field aligned with the rotation axis.
We neglect the field's backreaction on the geometry, so the metric remains
Kerr and the particles are subject to the Lorentz force. For a black hole
with zero net charge, we use the Wald electromagnetic potential~\cite{Wald:1974magnetic},
\begin{equation}
A_\mu=\frac{B}{2}\left(g_{\mu\phi}+2a g_{\mu t}\right),
\label{eq:wald_potential}
\end{equation}
where $B$ is the magnetic-field strength at infinity. With the units and
particle-mass normalization of the previous section, the motion is governed
by the Hamiltonian
\begin{equation}
H=\frac12 g^{\mu\nu}
\left(p_\mu-q_{\rm p}A_\mu\right)
\left(p_\nu-q_{\rm p}A_\nu\right)
=-\frac12,
\label{eq:charged_hamiltonian}
\end{equation}
where $q_{\rm p}$ is the particle charge and $p_\mu$ is the canonical momentum.
We denote the effective magnetic coupling by $b=q_{\rm p}B$. The explicit
model and equations of motion are given in Ref.~\cite{Cao:2026tpcd}.

Stationarity and axial symmetry give the conserved energy $E=-p_t$ and axial
angular momentum $L=p_\phi$. Fixing these quantities reduces the motion to
the radial--polar phase space. The magnetic coupling generally breaks Kerr
geodesic separability, allowing regular regions, resonant structures, and
chaotic regions to coexist. For two chaotic orbits, we use SWFR to locate
intervals near simple rational ratios and examine their Poincar\'e sections.

\subsection{Local structures near the 3/5 and 4/7 resonances in a single chaotic orbit}\label{sec:resonance_geometry}

We first consider an orbit with
\[
E=0.905,\qquad L=2,\qquad a=0.99,\qquad b=0.105.
\]
The initial conditions are
\[
r_0=4.6,\qquad \theta_0=\frac{\pi}{2},\qquad
\phi_0=0,\qquad p_{r0}=0.
\]
The mass-shell constraint gives the positive initial value
$p_{\theta0}\simeq1.480760952576273$. This orbit was
identified as chaotic by both TPCD and FLI in Ref.~\cite{Cao:2026tpcd}.
We integrate it with the eighth-order Runge--Kutta method, using a proper-time
step $h=0.1$ up to $\tau=10^7$, and calculate SWFR with $W=3000$. Following
Sec.~\ref{sec:periodic_structure}, we record $(r,p_r)$ whenever the orbit
crosses the equatorial plane with $\dot\theta>0$.

\begin{figure}[!t]
\centering
\includegraphics[width=0.95\textwidth]{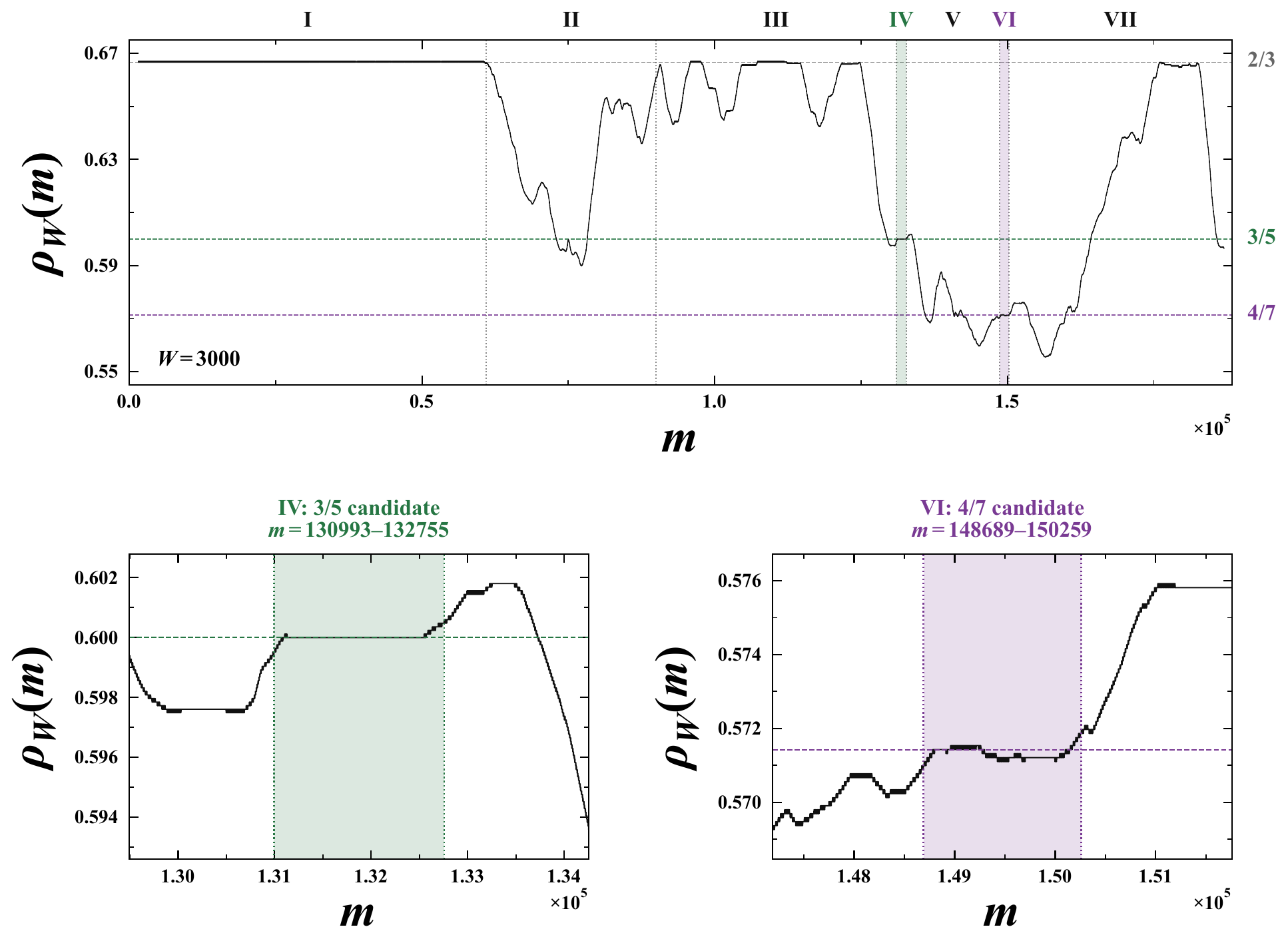}
\caption{SWFR for a chaotic charged-particle orbit with $E=0.905$,
$L=2$, $a=0.99$, $b=0.105$, and $r_0=4.6$, using $W=3000$.
Top: full evolution, with vertical lines separating stages I--VII.
Bottom: enlargements of the $3/5$ (left) and $4/7$ (right) candidate
intervals. Shading marks the selected intervals, and horizontal dashed
lines indicate the reference ratios. Here $m$ is the central radial event index.}
\label{fig:charged_swfr}
\end{figure}

The upper panel of Fig.~\ref{fig:charged_swfr} shows the full SWFR curve.
The ratio first stays close to $2/3$ for a long interval. It then varies
substantially, returning several times to the neighborhood of this value. Shorter stays
near $3/5$ and $4/7$ occur later. These changes are much larger than the
counting fluctuations around the fixed ratios in the regular benchmarks.
The lower panels of Fig.~\ref{fig:charged_swfr} show the intervals near
$3/5$ and $4/7$ in detail.

For each of the two target ratios, we require every window center in a
candidate interval to satisfy
\begin{equation}
\left|\rho_W(m_c)-\frac pq\right|\leq5\times10^{-4},
\label{eq:target_selection}
\end{equation}
and require a continuous span of at least $600$ in the central event index.
The two selected intervals are marked in the figure. Their central-index
endpoints and the corresponding proper times from the event table are
listed in Table~\ref{tab:charged_candidate_intervals}.

\begin{table}[!htbp]
\centering
\caption{Candidate intervals for the orbit in Fig.~\ref{fig:charged_swfr}.
The index endpoints are the first and last qualifying window centers.
Section points are selected over $[\tau_{m_a},\tau_{m_b})$. The displayed
times are rounded; the full-precision event times are used in the calculation.}
\label{tab:charged_candidate_intervals}
\begingroup
\renewcommand{\arraystretch}{1.15}
\begin{tabular}{ccccc}
\hline
Stage & Target ratio & $[m_a,m_b]$ & $[\tau_{m_a},\tau_{m_b})/10^6$ & Section points \\
\hline
IV & $3/5$ & $[130993,132755]$ & $[6.993721,7.086445)$ & 2937 \\
VI & $4/7$ & $[148689,150259]$ & $[7.907931,7.988539)$ & 2748 \\
\hline
\end{tabular}
\endgroup
\end{table}

The mean SWFR values are approximately $0.6000214$ and $0.5713323$,
respectively, close to the target ratios. Stage IV contains a flat portion
near $0.6$, while stage VI retains small fluctuations around $4/7$.
The criterion requires the ratio to remain within the specified tolerance;
it does not require an exactly horizontal curve.

We divide the full integration into seven consecutive stages, I--VII, using
the candidate-interval endpoints and two additional descriptive boundaries
at $m=61000$ and $m=90000$. Stage I starts at $\tau=0$, and stage VII ends
at $\tau=10^7$. All internal time boundaries come from the event table.
The stages are adjacent, and each section point belongs to exactly one
stage. They cover both candidate intervals and all motion before, between,
and after them. The boundaries organize the finite-time observations; they
do not specify exact dynamical transition times.

\begin{figure}[!t]
\centering
\includegraphics[width=\textwidth]{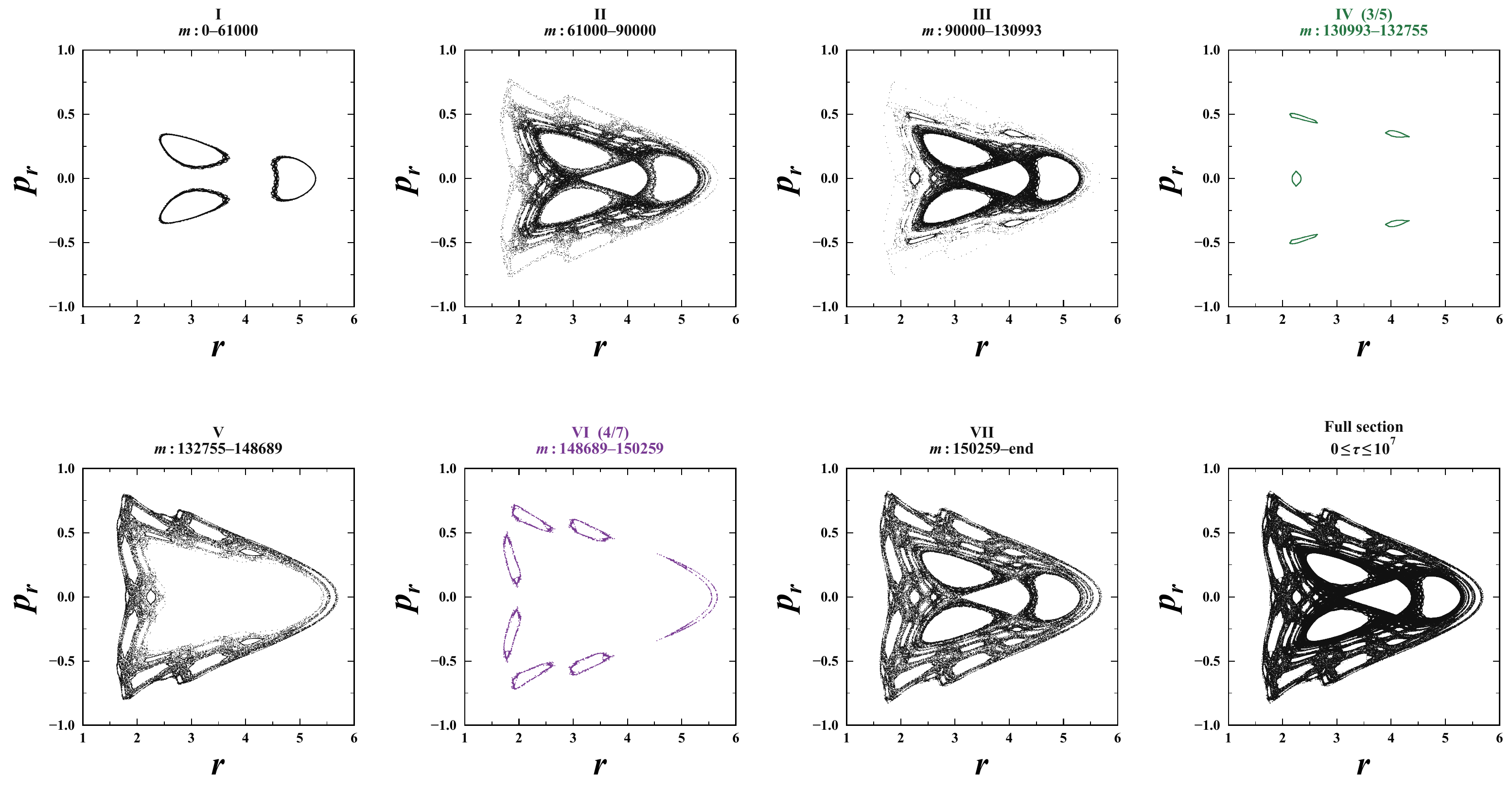}
\caption{Poincar\'e sections at $\theta=\pi/2$ with $\dot\theta>0$
for the seven stages in Fig.~\ref{fig:charged_swfr}. Stages IV (green)
and VI (purple) show fivefold and sevenfold structures near $3/5$ and
$4/7$, respectively. The lower-right panel is the full section over
$0\leq\tau\leq10^7$.}
\label{fig:charged_sections}
\end{figure}

Figure~\ref{fig:charged_sections} shows the sections for these stages.
In stage I, narrow branches surround three regions, consistent with the
long stay near $2/3$ in Fig.~\ref{fig:charged_swfr}. In stage II, SWFR
departs substantially from this plateau, and the section points spread
over a wider region. Stage III includes repeated returns toward $2/3$,
interrupted by changes in the ratio. Its accumulated section therefore
contains broad bands: repeated approaches to the same ratio do not
necessarily confine the points to a single narrow pattern.

In stage IV, the $2937$ section points form five separate, narrow
structures, consistent with the $3/5$ frequency relation. In stage V,
the ratio moves toward lower values and the points again spread over a
larger region. In stage VI, SWFR stays near $4/7$ and the $2748$ points
form seven main branches, some shaped as elongated arcs. This agrees
with the sevenfold structure discussed in Sec.~\ref{sec:periodic_structure}.
Stage VII contains further large changes in the ratio and a broader
distribution of section points.

The lower-right panel of Fig.~\ref{fig:charged_sections} combines all seven
stages into a section containing $296142$ points. Patterns from different
stages overlap, obscuring the fivefold and sevenfold structures of stages
IV and VI. Selecting the intervals with SWFR separates these structures
from the full accumulated section.

The same chaotic orbit thus approaches different resonant structures at
different times. Intervals where SWFR remains near $3/5$ and $4/7$ have
fivefold and sevenfold section geometries. This agreement supports
temporary residence or sticking near the respective resonances.
The point sets describe the chaotic orbit near these structures; they are
not the stable islands themselves and do not establish strictly periodic
motion during these intervals.

\subsection{Local structure near the 1/2 resonance in a second chaotic orbit}\label{sec:half_resonance}

For the second chaotic orbit, we keep $E=0.905$, $L=2$, and $a=0.99$, but
change the magnetic coupling to $b=0.057$ and the initial radius to
$r_0=1.8$. The other initial conditions are $\theta_0=\pi/2$, $\phi_0=0$,
and $p_{r0}=0$. The positive root of the mass-shell constraint is
$p_{\theta0}\simeq1.015289362248785$. Both TPCD
and FLI identified this orbit as chaotic in Ref.~\cite{Cao:2026tpcd}.
We use the same integration method, step size, and final time as above,
with $W=3000$.

\begin{figure}[!tp]
\centering
\includegraphics[width=0.95\textwidth]{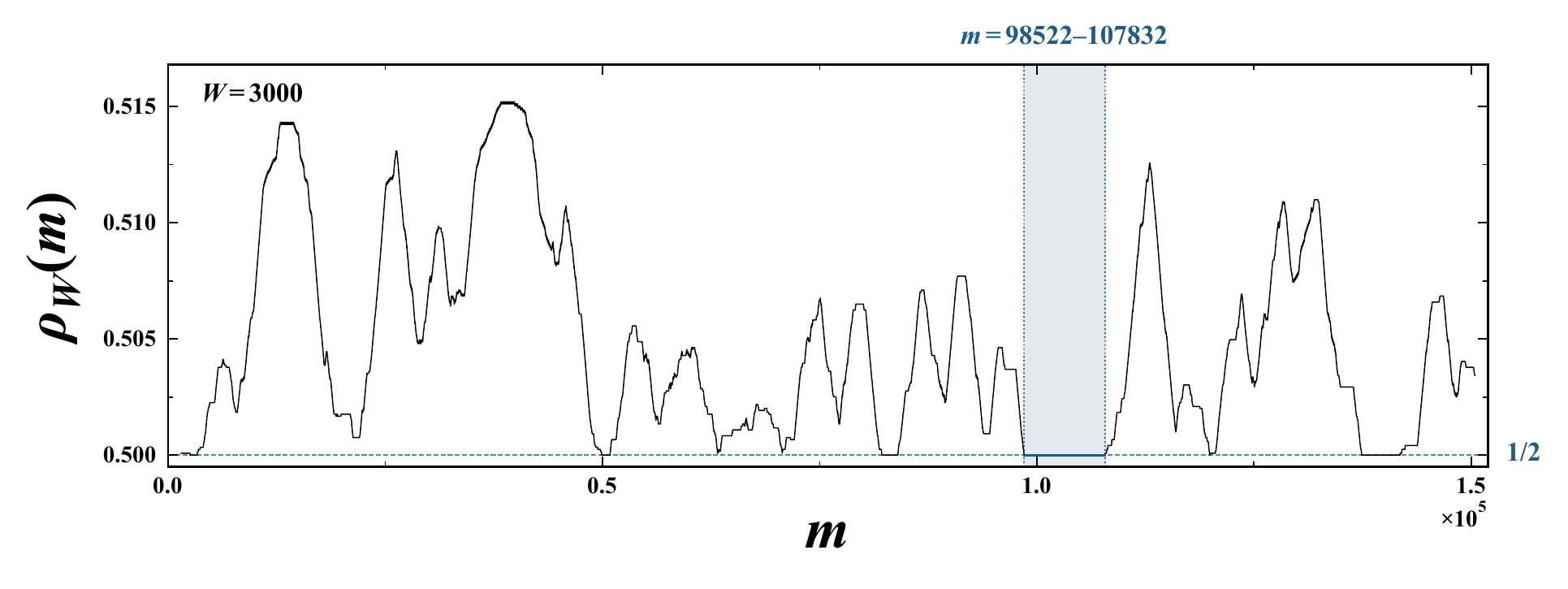}\par
(a)\par\vspace{0.7em}
\includegraphics[width=0.55\textwidth]{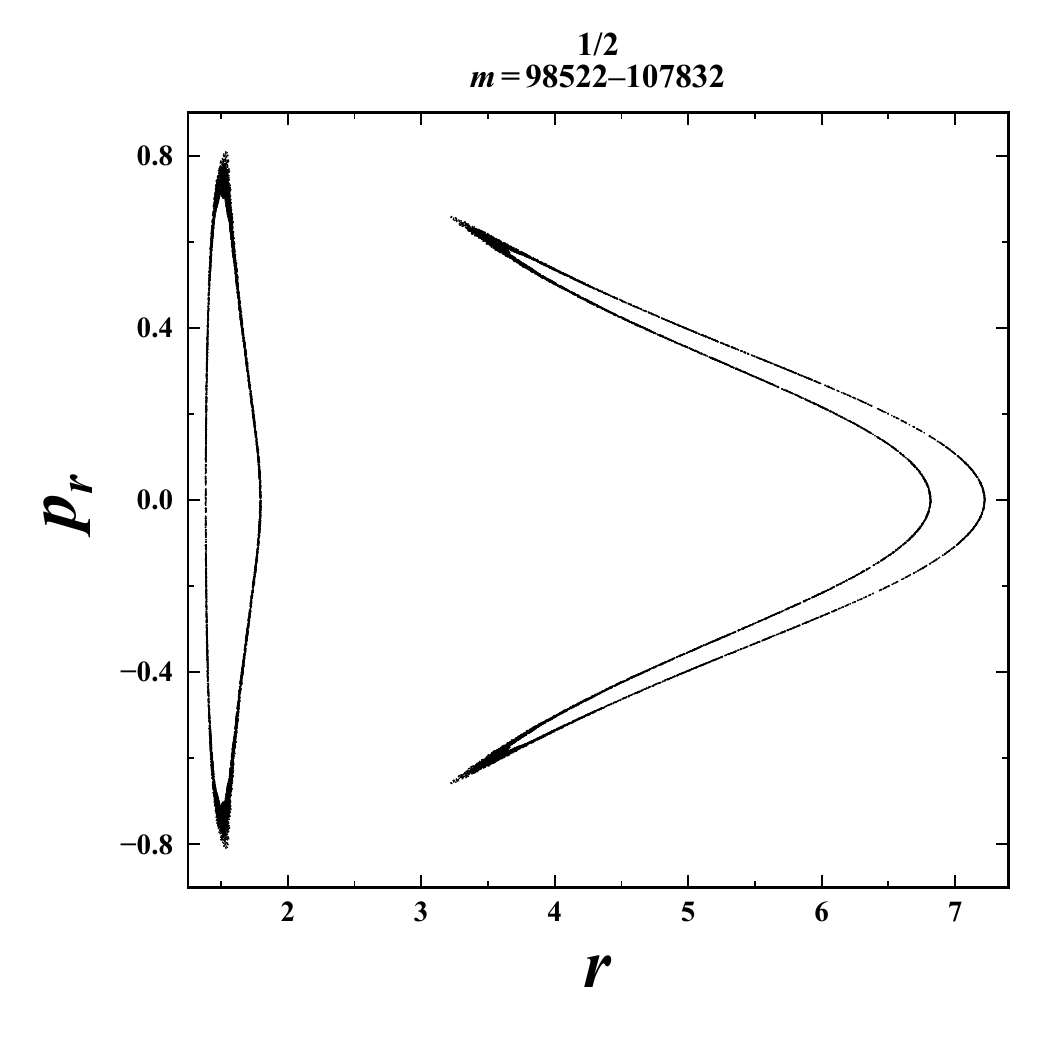}\par
(b)
\caption{SWFR and Poincar\'e section of a chaotic charged-particle
orbit with $E=0.905$, $L=2$, $a=0.99$, $b=0.057$, and $r_0=1.8$.
(a) SWFR for $W=3000$. The shaded band marks $m=98522$--$107832$,
and the dashed line indicates $1/2$.
(b) Section at $\theta=\pi/2$ with $\dot\theta>0$ over the corresponding
proper-time interval, showing a two-lobed structure.}
\label{fig:half_resonance}
\end{figure}

Figure~\ref{fig:half_resonance}(a) shows the full SWFR curve. The frequency
ratio changes repeatedly along the orbit and returns several times to the
neighborhood of $1/2$. The shaded interval,
\begin{equation}
m_c\in[98522,107832],
\label{eq:half_center_interval}
\end{equation}
is one of the longer plateaus, with every computed SWFR value equal to
$0.5$. The curve departs from this value on both sides of the plateau,
so the count relation holds for only part of the evolution.

The original event sequence shows two selected polar events in each
complete radial cycle within these windows. Each window of $3000$ radial
cycles therefore contains $6000$ polar events, giving
$\rho_W=3000/6000=1/2$.

Using the event table, we map the central-index interval to the proper-time
interval
\begin{equation}
\tau\in[6.489568859\times10^6,\;7.099476381\times10^6).
\label{eq:half_time_interval}
\end{equation}
This interval lasts approximately $6.0991\times10^5$ and contains $9310$
complete radial cycles. Applying the same section condition to all crossings
within it gives $18620$ section points.

The points in Fig.~\ref{fig:half_resonance}(b) form two lobes in the
$(r,p_r)$ plane. The branch at smaller radii is narrow and elongated; the
branch at larger radii spans a wider radial range. Most points lie along
narrow curve segments. Despite the different lobe shapes, the twofold
geometry is consistent with the return structure associated with $1/2$.
Every point in the selected time interval is retained, with no selection
by geometric appearance.

As in the fivefold and sevenfold examples, the SWFR plateau corresponds
to the section geometry over the same interval, here for a different
magnetic coupling and initial radius. The finite-window count ratio is
exactly $1/2$. Together with the two-lobed section, it supports a
period of residence near the $1/2$ resonant structure, but does not establish
a periodic orbit.

\section{Local resonance structure of Schwarzschild--Melvin photon orbits}\label{sec:melvin}

We also apply SWFR to photon orbits in Schwarzschild--Melvin spacetime.
This static, axisymmetric solution of the Einstein--Maxwell equations
includes the backreaction of the magnetic field on the geometry~\cite{Ernst:1976magnetic}. Photons
follow its null geodesics, allowing us to test the event-count method in
another nonintegrable relativistic system. The model and its null geodesic
equations are described in Ref.~\cite{LimaJunior:2021shadows}.

In units with $G=c=M=1$, the photon motion is governed by the Hamiltonian
\begin{equation}
H=\frac{1}{2\Lambda^2}
\left(-\frac{E^2}{f}+fp_r^2+\frac{p_\theta^2}{r^2}\right)
+\frac{\Lambda^2L^2}{2r^2\sin^2\theta}=0,
\label{eq:melvin_hamiltonian}
\end{equation}
where
\begin{equation}
f=1-\frac{2}{r},\qquad
\Lambda=1+\frac{B^2r^2\sin^2\theta}{4}.
\label{eq:melvin_functions}
\end{equation}
Here $B$ is the magnetic-field parameter, while $E=-p_t$ and $L=p_\phi$
are the conserved quantities associated with time translation and axial
symmetry. The photon orbit is parametrized by an affine parameter $\lambda$.
The turning-event definitions and sliding-window calculation remain the
same as in the preceding sections.

We take
\[
E=0.9043555924130631,\qquad L=4,\qquad B=0.18,
\]
with initial conditions
\[
r_0=3.953150601840506,\qquad \theta_0=\frac{\pi}{2},\qquad
\phi_0=0,\qquad p_{r0}=0.
\]
The null constraint $H=0$ gives the positive initial value
$p_{\theta0}\simeq0.3085906231811401$. We use the same eighth-order
Runge--Kutta method, with an affine-parameter step $h=0.1$, and integrate
up to $\lambda=10^7$. The window width is again $W=3000$.

\begin{figure}[!tp]
\centering
\includegraphics[width=0.95\textwidth]{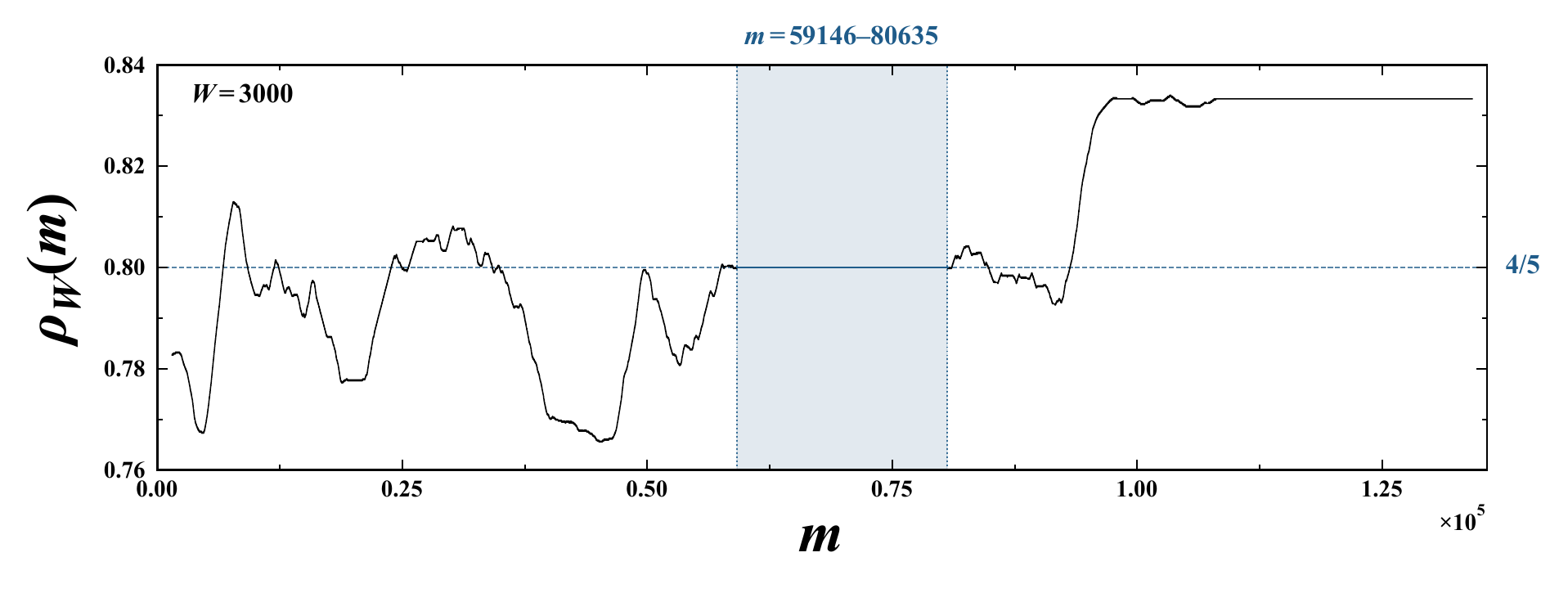}\par
(a)\par\vspace{0.7em}
\includegraphics[width=0.55\textwidth]{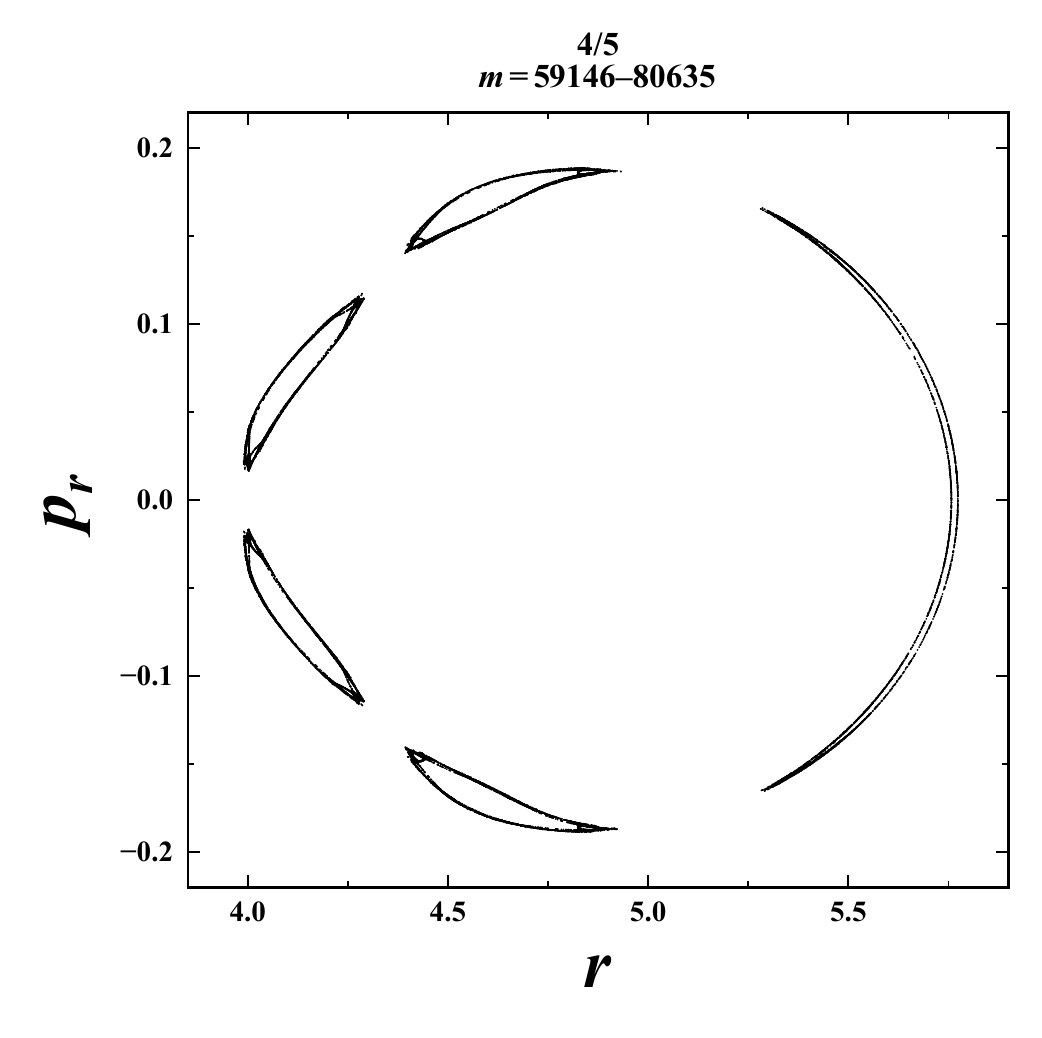}\par
(b)
\caption{SWFR and Poincar\'e section of the Schwarzschild--Melvin
photon orbit for the parameters and initial conditions given in
Sec.~\ref{sec:melvin}. (a) SWFR for $W=3000$. The shaded band marks
$m=59146$--$80635$, and the dashed line indicates $4/5$.
(b) Section at $\theta=\pi/2$ with $d\theta/d\lambda>0$ over the
corresponding affine-parameter interval, showing a fivefold structure.}
\label{fig:melvin_resonance}
\end{figure}

Figure~\ref{fig:melvin_resonance}(a) shows the full SWFR curve. The ratio
varies substantially early in the integration, then forms a long plateau
near $4/5$. After leaving the plateau, it changes again and reaches higher values
toward the end of the integration.

Following the continuous-interval search used in the previous section,
we require
\begin{equation}
\left|\rho_W(m_c)-\frac45\right|\leq10^{-4}
\label{eq:melvin_ratio_band}
\end{equation}
at every window center, with a span of at least $600$ in the central event
index. The tolerance is tighter than the $5\times10^{-4}$ used in
Sec.~\ref{sec:resonance_geometry}. As shown below, every computed SWFR
value in the selected interval equals $0.8$ exactly, so the result does not
depend on this choice; we adopt the stricter value as a conservative
criterion. This gives the shaded interval
\begin{equation}
m_c\in[59146,80635].
\label{eq:melvin_center_interval}
\end{equation}
Every computed SWFR value in this interval is $0.8$. Each window of
$3000$ radial cycles contains $3750$ selected polar events, giving
$\rho_W=3000/3750=4/5$.

Using the event table, we map this central-index interval to
\begin{equation}
\lambda\in[4.393624518\times10^6,\;5.980171727\times10^6).
\label{eq:melvin_affine_interval}
\end{equation}
Its affine-parameter length is approximately $1.58655\times10^6$, and it
contains $21489$ complete radial cycles. We record every crossing at
$\theta=\pi/2$ with $d\theta/d\lambda>0$ within this interval, obtaining
$26861$ section points. Their distribution in the $(r,p_r)$ plane is shown
in Fig.~\ref{fig:melvin_resonance}(b).

\FloatBarrier

The section points form five separated, elongated groups. Four lie at
smaller and intermediate radii; the fifth forms a longer arc at
larger radii. Despite their different shapes and sizes, the groups
have a fivefold geometry consistent with the period-five return structure
associated with the $4/5$ frequency relation. The section includes every
point recorded in the selected interval, without further selection by
geometric appearance.

The $4/5$ plateau and fivefold section structure in
Fig.~\ref{fig:melvin_resonance} are consistent with temporary residence near
the $4/5$ resonance. This extends the charged-particle results to null
geodesic motion in Schwarzschild--Melvin spacetime.

\section{Summary and discussion}\label{sec:conclusions}

We constructed the sliding-window frequency ratio (SWFR) from radial and
polar turning events to follow local frequency relations along individual
orbits. The method counts polar events within a window containing a fixed
number of radial cycles. The recorded event indices and occurrence times
link selected SWFR intervals to their Poincar\'e sections, preserving the
time information needed to follow local structures along the orbit.

The integrable Kerr tests illustrate both the frequency interpretation and
the effect of window width. For the quasiperiodic orbit, SWFR fluctuates
around the theoretical ratio $1/\sqrt{2}$, with smaller counting fluctuations
in larger windows. For periodic radial--polar motion with ratio $4/5$, SWFR
can remain constant when the window spans an integer number of repeating
units, although small windows can produce discrete fluctuations. These
results agree with the finite-window analysis: when the count deviation is
uniformly bounded, the counting-error bound for the ratio decreases as
$O(W^{-1})$. Interpreting changes in SWFR therefore requires distinguishing
finite-count effects from changes in the local oscillation relation.

For charged particles in Kerr spacetime with an external magnetic field,
SWFR identified candidate intervals near $3/5$ and $4/7$ along one chaotic
orbit. Their sections show fivefold and sevenfold local structures, respectively.
Dividing the evolution into seven consecutive stages reveals how these
structures alternate with more broadly distributed section points.
Separating the stages makes local features visible that are obscured by
overlap in the long-time section. With a different magnetic coupling and
initial radius, a second chaotic orbit exhibits a plateau near $1/2$ and
a two-lobed section during that interval.

For the Schwarzschild--Melvin photon orbit, the $4/5$ plateau corresponds
to a fivefold section structure over the selected affine-parameter interval.
Here the magnetic field affects the geometry through its own gravity, and
photons follow null geodesics.

The examples are consistent with globally chaotic orbits spending finite
intervals near different resonant structures and exploring broader regions
of phase space between these stays. SWFR locates candidate intervals through their
frequency relations, and the sections supply geometric evidence for this
interpretation. The method identifies local motion in resonant neighborhoods;
the organized point sets observed over finite intervals should not be
equated with the stable resonant islands themselves.

The selected events must correspond to complete oscillation cycles. The
window width must balance counting accuracy and time resolution: larger
windows reduce counting fluctuations but can mix neighboring stages and
smooth shorter changes. We used $W=3000$ in all nonintegrable examples;
this is not a universal choice. For other orbits, the window should be
chosen according to the oscillation scales, the duration of candidate
stages, and their Poincar\'e sections.

SWFR requires only turning events and their occurrence times, without
a spectral decomposition of the orbital signal. Once the events are
recorded, different windows and candidate frequency ratios can be examined
in post-processing.
Combining these counts with sections from the same intervals can reveal
local resonant patterns that overlap in the long-time section.

\section*{Acknowledgments}
This work is supported by the National Natural Science
Foundation of China (NSFC) under Grant nos. 12235019 and 12275106.


\begin{thebibliography}{99}

\bibitem{Carter:1968rr}
B.~Carter,
``Global structure of the Kerr family of gravitational fields,''
Phys. Rev. \textbf{174}, 1559-1571 (1968)
doi:10.1103/PhysRev.174.1559

\bibitem{Schmidt:2002frequencies}
W.~Schmidt,
``Celestial mechanics in Kerr spacetime,''
Class. Quantum Grav. \textbf{19}, 2743--2764 (2002),
\href{https://doi.org/10.1088/0264-9381/19/10/314}{doi:10.1088/0264-9381/19/10/314}.

\bibitem{Brink:2015resonances}
J.~Brink, M.~Geyer, and T.~Hinderer,
``Orbital Resonances Around Black Holes,''
Phys. Rev. Lett. \textbf{114}, 081102 (2015),
\href{https://doi.org/10.1103/PhysRevLett.114.081102}{doi:10.1103/PhysRevLett.114.081102}.

\bibitem{Mukherjee:2023resonance}
S.~Mukherjee, O.~Kop\'a\v{c}ek, and G.~Lukes-Gerakopoulos,
``Resonance crossing of a charged body in a magnetized Kerr background:
An analog of extreme mass ratio inspiral,''
Phys. Rev. D \textbf{107}, 064005 (2023),
\href{https://doi.org/10.1103/PhysRevD.107.064005}{doi:10.1103/PhysRevD.107.064005}.

\bibitem{Contopoulos:2010stickiness}
G.~Contopoulos and M.~Harsoula,
``Stickiness effects in chaos,''
Celest. Mech. Dyn. Astron. \textbf{107}, 77--92 (2010),
\href{https://doi.org/10.1007/s10569-010-9282-6}{doi:10.1007/s10569-010-9282-6}.

\bibitem{Laskar:1993frequency}
J.~Laskar,
``Frequency analysis for multi-dimensional systems. Global dynamics and diffusion,''
Physica D \textbf{67}, 257--281 (1993),
\href{https://doi.org/10.1016/0167-2789(93)90210-R}{doi:10.1016/0167-2789(93)90210-R}.

\bibitem{Chandre:2003timefrequency}
C.~Chandre, S.~Wiggins, and T.~Uzer,
``Time--frequency analysis of chaotic systems,''
Physica D \textbf{181}, 171--196 (2003),
\href{https://doi.org/10.1016/S0167-2789(03)00117-9}{doi:10.1016/S0167-2789(03)00117-9}.

\bibitem{LukesGerakopoulos:2010signature}
G.~Lukes-Gerakopoulos, T.~A.~Apostolatos, and G.~Contopoulos,
``Observable signature of a background deviating from the Kerr metric,''
Phys. Rev. D \textbf{81}, 124005 (2010),
\href{https://doi.org/10.1103/PhysRevD.81.124005}{doi:10.1103/PhysRevD.81.124005}.

\bibitem{LukesGerakopoulos:2012zipoy}
G.~Lukes-Gerakopoulos,
``Nonintegrability of the Zipoy--Voorhees metric,''
Phys. Rev. D \textbf{86}, 044013 (2012),
\href{https://doi.org/10.1103/PhysRevD.86.044013}{doi:10.1103/PhysRevD.86.044013}.

\bibitem{Zelenka:2020spin}
O.~Zelenka, G.~Lukes-Gerakopoulos, V.~Witzany, and O.~Kop\'a\v{c}ek,
``Growth of resonances and chaos for a spinning test particle in the Schwarzschild background,''
Phys. Rev. D \textbf{101}, 024037 (2020),
\href{https://doi.org/10.1103/PhysRevD.101.024037}{doi:10.1103/PhysRevD.101.024037}.

\bibitem{Destounis:2020symmetry}
K.~Destounis, A.~G.~Suvorov, and K.~D.~Kokkotas,
``Testing spacetime symmetry through gravitational waves from extreme-mass-ratio inspirals,''
Phys. Rev. D \textbf{102}, 064041 (2020),
\href{https://doi.org/10.1103/PhysRevD.102.064041}{doi:10.1103/PhysRevD.102.064041}.

\bibitem{Destounis:2021glitches}
K.~Destounis and K.~D.~Kokkotas,
``Gravitational-wave glitches: Resonant islands and frequency jumps in nonintegrable extreme-mass-ratio inspirals,''
Phys. Rev. D \textbf{104}, 064023 (2021),
\href{https://doi.org/10.1103/PhysRevD.104.064023}{doi:10.1103/PhysRevD.104.064023}.

\bibitem{Deich:2022quadratic}
A.~Deich, A.~C\'ardenas-Avenda\~no, and N.~Yunes,
``Chaos in quadratic gravity,''
Phys. Rev. D \textbf{106}, 024040 (2022),
\href{https://doi.org/10.1103/PhysRevD.106.024040}{doi:10.1103/PhysRevD.106.024040}.

\bibitem{EduarteRojas:2022kerrlike}
A.~Eduarte-Rojas, F.~Frutos-Alfaro, R.~Carboni, and D.~Alvarado,
``Chaotic behavior of geodesics in Kerr-like spacetime,''
Phys. Rev. D \textbf{106}, 064015 (2022),
\href{https://doi.org/10.1103/PhysRevD.106.064015}{doi:10.1103/PhysRevD.106.064015}.

\bibitem{Chen:2022eob}
C.-Y.~Chen, F.-L.~Lin, and A.~Patel,
``Resonant islands of effective-one-body dynamics,''
Phys. Rev. D \textbf{106}, 084064 (2022),
\href{https://doi.org/10.1103/PhysRevD.106.084064}{doi:10.1103/PhysRevD.106.084064}.

\bibitem{Destounis:2023boson}
K.~Destounis, F.~Angeloni, M.~Vaglio, and P.~Pani,
``Extreme-mass-ratio inspirals into rotating boson stars: Nonintegrability, chaos, and transient resonances,''
Phys. Rev. D \textbf{108}, 084062 (2023),
\href{https://doi.org/10.1103/PhysRevD.108.084062}{doi:10.1103/PhysRevD.108.084062}.

\bibitem{Shahzadi:2023hartlethorne}
M.~Shahzadi, M.~Kolo\v{s}, R.~Saleem, Y.~Habib, and A.~Eduarte-Rojas,
``Structure-preserving numerical simulations of test particle dynamics around slowly rotating neutron stars within the Hartle--Thorne approach,''
Phys. Rev. D \textbf{108}, 103006 (2023),
\href{https://doi.org/10.1103/PhysRevD.108.103006}{doi:10.1103/PhysRevD.108.103006}.

\bibitem{Destounis:2026environment}
K.~Destounis and P.~G.~S.~Fernandes,
``Environmentally induced chaos: Extreme-mass-ratio systems of rotating black holes in astrophysical environments,''
Phys. Rev. D \textbf{113}, 044040 (2026),
\href{https://doi.org/10.1103/slvg-d6mh}{doi:10.1103/slvg-d6mh}.

\bibitem{Strateny:2026matter}
M.~Straten\'y, G.~Lukes-Gerakopoulos, and O.~Zelenka,
``Extreme mass ratio inspirals into black holes surrounded by matter: Resonance crossings,''
Phys. Rev. D \textbf{113}, 044054 (2026),
\href{https://doi.org/10.1103/qxx6-vvvk}{doi:10.1103/qxx6-vvvk}.

\bibitem{Yuan:2026spin}
D.-D.~Yuan, J.-G.~Jiao, Y.-Q.~Lei, J.-X.~Shi, J.-Q.~Lai, C.~Shao, and Y.~Tian,
``Astrophysically realistic secondary spins trigger chaos in Schwarzschild spacetime and discernible gravitational wave signatures,''
Phys. Rev. D \textbf{114}, 064043 (2026),
\href{https://doi.org/10.1103/zjkr-hvvz}{doi:10.1103/zjkr-hvvz}.

\bibitem{Szezech:2013ftrn}
J.~D.~Szezech, Jr., A.~B.~Schelin, I.~L.~Caldas, S.~R.~Lopes,
P.~J.~Morrison, and R.~L.~Viana,
``Finite-time rotation number: A fast indicator for chaotic dynamical structures,''
Phys. Lett. A \textbf{377}, 452--456 (2013),
\href{https://doi.org/10.1016/j.physleta.2012.12.013}{doi:10.1016/j.physleta.2012.12.013}.

\bibitem{Santos:2019sticky}
M.~S.~Santos, M.~Mugnaine, J.~D.~Szezech, Jr., A.~M.~Batista,
I.~L.~Caldas, and R.~L.~Viana,
``Using rotation number to detect sticky orbits in Hamiltonian systems,''
Chaos \textbf{29}, 043125 (2019),
\href{https://doi.org/10.1063/1.5078533}{doi:10.1063/1.5078533}.

\bibitem{Apostolatos:2009nonKerr}
T.~A.~Apostolatos, G.~Lukes-Gerakopoulos, and G.~Contopoulos,
``How to Observe a Non-Kerr Spacetime Using Gravitational Waves,''
Phys. Rev. Lett. \textbf{103}, 111101 (2009),
\href{https://doi.org/10.1103/PhysRevLett.103.111101}{doi:10.1103/PhysRevLett.103.111101}.

\bibitem{Semerak:2012geodesics}
O.~Semer\'ak and P.~Sukov\'a,
``Free motion around black holes with discs or rings:
between integrability and chaos--II,''
Mon. Not. R. Astron. Soc. \textbf{425}, 2455--2476 (2012),
\href{https://doi.org/10.1111/j.1365-2966.2012.21630.x}{doi:10.1111/j.1365-2966.2012.21630.x}.

\bibitem{Yurin:2024zerocrossing}
D.~Yurin, M.~Makukov, D.~Kuvatova, A.~Gluchshenko, and C.~Omarov,
``Applying zero-crossing method for frequency map analysis of dynamical systems,''
Mon. Not. R. Astron. Soc. \textbf{531}, 1483--1495 (2024),
\href{https://doi.org/10.1093/mnras/stae1216}{doi:10.1093/mnras/stae1216}.

\bibitem{LimaJunior:2021shadows}
H.~C.~D.~Lima Junior, P.~V.~P.~Cunha, C.~A.~R.~Herdeiro, and L.~C.~B.~Crispino,
``Shadows and lensing of black holes immersed in strong magnetic fields,''
Phys. Rev. D \textbf{104}, 044018 (2021),
\href{https://doi.org/10.1103/PhysRevD.104.044018}{doi:10.1103/PhysRevD.104.044018}.

\bibitem{Cao:2026tpcd}
W.~Cao, Y.~Wang, and H.~Zhang,
``Turning-Point Count Discrepancy as a Diagnostic of Relativistic Orbital Chaos,''
\href{https://arxiv.org/abs/2609.28914}{arXiv:2609.28914 [gr-qc]} (2026).

\bibitem{Wald:1974magnetic}
R.~M.~Wald,
``Black hole in a uniform magnetic field,''
Phys. Rev. D \textbf{10}, 1680--1685 (1974),
\href{https://doi.org/10.1103/PhysRevD.10.1680}{doi:10.1103/PhysRevD.10.1680}.

\bibitem{Ernst:1976magnetic}
F.~J.~Ernst,
``Black holes in a magnetic universe,''
J. Math. Phys. \textbf{17}, 54--56 (1976),
\href{https://doi.org/10.1063/1.522781}{doi:10.1063/1.522781}.

\end{thebibliography}
\end{document}